\documentclass[showpacs,twocolumn,aps,floatfix]{revtex4-1}%
\usepackage{amssymb}
\usepackage{amsfonts}
\usepackage{amsmath}
\usepackage{graphicx}%
\providecommand{\U}[1]{\protect\rule{.1in}{.1in}}

\begin{document}
\preprint{cond-mat/}
\title[Short title for running header]{Dynamics of Energy Fluctuations in Equilibrating and Driven-Dissipative Systems}
\author{Guy Bunin}
\affiliation{Department of Physics, Technion - Israel Institute of Technology, Haifa 32000}
\author{Yariv Kafri}
\affiliation{Department of Physics, Technion - Israel Institute of Technology, Haifa 32000}

\begin{abstract}
When two isolated system are brought in contact, they relax to equilibrium via
energy exchange. In another setting, when one of the systems is driven and the
other is large, the first system reaches a steady-state which is not described
by the Gibbs distribution. Here, we derive expressions for the size of energy
fluctuations as a function of time in both settings, assuming that the process
is composed of many small steps of energy exchange. In both cases the results
depend only on the average energy flows in the system, independent of any
other microscopic detail. In the steady-state we also derive an expression
relating three key properties: the relaxation time of the system, the energy
injection rate, and the size of the fluctuations.

\end{abstract}
\volumeyear{year}
\volumenumber{number}
\issuenumber{number}
\eid{identifier}
\startpage{1}
\endpage{2}
\maketitle

\begin{figure}
[ptb]
\begin{center}
\includegraphics[
height=1.7729in,
width=3.3961in
]%
{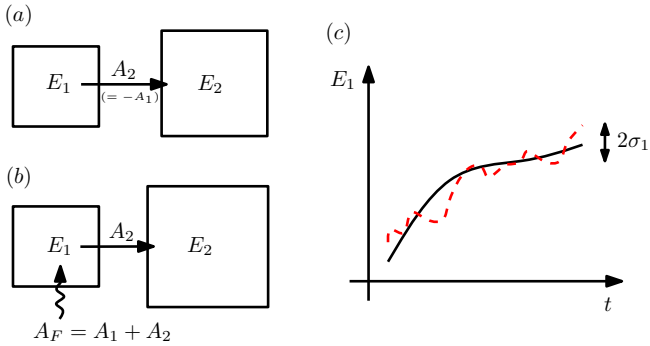}%
\caption{The energy fluctuations in the two set-ups. (a) Exchange of energy
between two systems. (b) A system driven by an external force and attached to
a bath. (c) A typical evolution of $E_{1}\left(  t\right)  $ (dashed line)
fluctuates around the average $\left\langle E_{1}\right\rangle \left(
t\right)  $ (solid line). Eq. (\ref{eq:steady_state_approach}) relates the
size of these fluctuations to the average $\left\langle E_{1}\right\rangle
\left(  t\right)  $.}%
\label{fig:settings}%
\end{center}
\end{figure}

In this paper we consider two closely related non-equilibrium problems. In the
first problem, two systems which are coupled to each other but isolated
otherwise, are allowed to exchange energy, see Fig. \ref{fig:settings}(a). The
systems start with arbitrary initial energies and eventually reach
equilibrium. It is natural to ask: How do the initial energies evolve in time
as the two systems approach equilibrium? For example, one might imagine
measuring the energy of a tea cup as it cools, or the equilibration of a
mesoscopic system of two atomic gases, initially prepared at two different
temperatures. In the second problem, one of the two systems is also driven by
an external protocol, see Fig. \ref{fig:settings}(b). This is achieved, for
example, by applying a time-varying field which repeatedly returns to its
initial form. When the second system is much larger than the first, it acts as
a dissipative bath, and the first system eventually settles to a
non-equilibrium steady-state. This scenario serves as a generic model for
driven-dissipative systems, which describe a broad range of phenomena
\cite{granular_gases,SL_PRL,DD_refs}. Here, one can ask how the first system
reaches a steady-state, and what are the properties of this non-Gibbsian
steady-state.%

As the dynamics of a system are affected by the detailed microscopic state,
repeating the same experiment will lead to different outcomes. Specifically, a
measurement of the energy as a function of time will yield different results,
see Fig. \ref{fig:settings}(c). The variations between experiments might
average-out in large, thermodynamic systems, or when the driving protocol
applied is quasi-static. However, they can be significant when the drive is
not quasi-static and in small or mesoscopic systems, which are of current
experimental interest \cite{cold_atoms,trapped_ions,nuc_spins}. Here we
quantify these energy fluctuations by studying the variance of the energy
measurements in repeated experiments. The dependence of these fluctuations on
the dynamics makes general statements scarce, and one typically has to resort
to the study of specific models.

In this Letter we show that when the changes in energy are small and slow (but
still irreversible), general statements about the energy fluctuations can be
made. The results are insensitive to almost all microscopic details of the
systems, depending only on the average energy flows from the drive to the
system and between the systems as a function of time, and on the density of
states. We stress that the assumptions made do not imply that the combined
system (composed of systems 1 and 2)\ is close to equilibrium, but only that
each of the systems separately is close to equilibrium within its energy
shell. Our main results are: (1) Eqs. (\ref{eq:equilibrium_approach}) and
(\ref{eq:steady_state_approach}), which quantify the variance of the energy
fluctuations as the system approaches its steady-state (which is equilibrium
when no drive is present). (2) Eq. (\ref{eq:steady_state_tau}), which relates
three main quantities at the steady-state: the variance of the energy
fluctuations, the average rate of energy flow through the system, and the
relaxation time of energy fluctuations. The validity of the results is
illustrated in a system of colliding hard spheres.

To derive them we consider the evolution of the energies in the $\left(
E_{1},E_{2}\right)  $ plane, where $E_{1},E_{2}$ are the energies of systems 1
and 2 respectively. Consider a series of small changes in the energies, each
taking place over a time interval $\Delta t$. We assume that $\tau_{R}%
\ll\Delta t$, where $\tau_{R}$ is the relaxation time of each of the isolated
systems separately. When this time scale separation holds, the statistics of
the energy changes $\Delta E_{1},\Delta E_{2}$ during the time interval from
$t$ to $t+\Delta t$ depend only on the energies $\left(  E_{1},E_{2}\right)  $
at time $t$. The time evolution of the probability distribution $P_{12}\left(
E_{1},E_{2}\right)  $ is then governed by a Fokker-Planck equation%
\begin{equation}
\partial_{t}P_{12}=-\sum_{i=1}^{2}\partial_{E_{i}}\left(  A_{i}P_{12}\right)
+\frac{1}{2}\sum_{i,j=1}^{2}\partial_{E_{i}}\partial_{E_{j}}\left(
B_{ij}P_{12}\right)  \label{eq.FP}%
\end{equation}
where $A_{1},A_{2},B_{11},B_{12},B_{21},B_{22}$ are all functions of $\left(
E_{1},E_{2}\right)  \,$, and $B_{12}=B_{21}$. These function are related to
the first two moments of the changes in $E_{1},E_{2}$ during a \emph{short}
time $\Delta t$ \cite{gardiner,cumulant_comment}:%
\[
A_{i}=\frac{\left\langle \Delta E_{i}\right\rangle }{\Delta t}\ ;\ B_{ij}%
=\frac{\left\langle \Delta E_{i}\Delta E_{j}\right\rangle }{\Delta t}\ .
\]
The equation is valid when higher cumulants, e.g. $\left\langle \Delta
E_{i}\Delta E_{j}\Delta E_{k}\right\rangle _{c}/\Delta t$, are small compared
to the $A_{i}$ and $B_{ij}$ functions.

In both scenarios, of equilibrating systems and driven-dissipative systems,
one can take the $A_{i}$ and $B_{ij}$ functions to depend on $E_{1}$ only. For
equilibrating systems, this is possible when the initial total energy
$E_{total}=E_{1}+E_{2}$\ is fixed, so that $E_{2}$ can be considered to\ be a
function of $E_{1}$. In the case of driven-dissipative systems, $E_{2}$ drops
completely from the equations when we take system 2 to be much larger than
system 1. As shown below, this is because system 2 acts as a thermal bath
whose properties are insensitive to the changes in $E_{2}$. It is then more
convenient to work with the marginal probability distribution of $E_{1}$
alone: $P\left(  E_{1}\right)  \equiv\int dE_{2}P_{12}\left(  E_{1}%
,E_{2}\right)  $. Integrating Eq. (\ref{eq.FP}) over $E_{2}$ we find%
\begin{equation}
\partial_{t}P=-\partial_{E_{1}}\left(  A_{1}P\right)  +\frac{1}{2}%
\partial_{E_{1}}^{2}\left(  B_{11}P\right)  \ , \label{eq:FP1}%
\end{equation}
\newline using $P_{12}\left(  E_{2}\rightarrow\pm\infty\right)  =0$. \ While
only the functions $A_{1}$ and $B_{11}$ appear in this equation, the
interaction with system 2 still affects the energy of system 1, via the forms
of the functions $A_{1},B_{11}$. This is contained in the relation which is
derived below%
\begin{equation}
2A_{1}-2\beta_{2}/\beta_{1}A_{F}=\left(  \beta_{1}-\beta_{2}\right)  B_{11}\ ,
\label{eq:full_relation}%
\end{equation}
where $A_{F}\equiv A_{1}+A_{2}=\partial_{t}\left\langle E_{total}\right\rangle
$ is the rate of energy injected into the system by the drive. The inverse
temperatures are defined by $\beta_{1}\left(  E_{1}\right)  =\partial_{E_{1}%
}S_{1}\left(  E_{1}\right)  $, and $\beta_{2}\left(  E_{2}\right)
=\partial_{E_{2}}S_{2}\left(  E_{2}\right)  $, where $S_{1,2}$ are the
(microcanonical) entropies of systems 1,2 respectively. $\beta_{1}$ and
$\beta_{2}$ are well-defined functions, depending only on the density of
states of the system, and unrelated to the driving mechanism and the
interaction between the systems. Moreover, $\beta_{1}$ can be very different
from $\beta_{2}$. Eq. (\ref{eq:full_relation}) is ultimately based on
Liouville's equation, or the unitarity of the dynamics in quantum cases. In
the driven case we also assume that the energy flow from the drive and between
the systems are statistically independent processes, see discussion below. Eq.
(\ref{eq:full_relation}) is exact up to corrections of order $1/N$, where $N$
is the number of degrees of freedom of the smaller of the two systems. In the
case of equilibrating systems $A_{F}=0$, and the relation Eq.
(\ref{eq:full_relation}) reduces to $2A_{1}=\left(  \beta_{1}-\beta
_{2}\right)  B_{11}$. Here, as expected, on average energy flows from high to
low temperatures.

The drive is implemented by varying the Hamiltonian of system 1 in time (e.g.,
by applying a time-varying external field). We consider drives where the
Hamiltonian repeatedly returns to its original form (i.e., an oscillating
field). At the steady-state, when the Hamiltonian is changed adiabatically,
returning to the original form leaves the energy of the combined system
unchanged. Thus, the changes in the energy will only be due to irreversible effects.

Before deriving Eq. (\ref{eq:full_relation}) we consider several of its
consequences in the two scenarios, of equilibrating and driven-dissipative
systems. Wherever possible, we present the results in a unified way where the
case of equilibrating systems is obtained by setting $A_{F}=0$.

\emph{Approach to steady-state - }We start by considering the approach of the
combined system (composed of systems 1 and 2) to its steady-state. If no
driving is present (scenario 1), this steady-state is thermal equilibrium. We
derive an expression for the evolution of the variance $\sigma_{1}%
^{2}=\left\langle E_{1}^{2}\right\rangle -\left\langle E_{1}\right\rangle
^{2}$ during the entire equilibration process. Proceeding similarly to
\cite{NatPhys}, we take the first two moments with respect to $E_{1}$ of Eq.
(\ref{eq:FP1})
\begin{align}
\partial_{t}\left\langle E_{1}\right\rangle  &  =\left\langle A_{1}%
\right\rangle \ ,\nonumber\\
\partial_{t}\sigma_{1}^{2} &  =\left\langle B_{11}\right\rangle +2\left(
\left\langle A_{1}E_{1}\right\rangle -\left\langle A_{1}\right\rangle
\left\langle E_{1}\right\rangle \right)  \ .\label{eq:ODEs}%
\end{align}
If the distribution is narrow enough (valid up to $1/N$ corrections, see
discussion after Eq. (\ref{eq:steady_state_approach})), $\left\langle
A_{1}\right\rangle $ can be assumed to depend on $\left\langle E_{1}%
\right\rangle $ alone, and the change in $\left\langle E_{1}\right\rangle $
will be monotonic. Combining the two equalities in Eq. (\ref{eq:ODEs}) and
linearizing $A_{1}$ within the width of the probability distribution, we find
\begin{equation}
\frac{\partial\sigma_{1}^{2}}{\partial\left\langle E_{1}\right\rangle
}=2Z\left(  \left\langle E_{1}\right\rangle \right)  +2\frac{\partial_{E_{1}%
}A_{1}\left(  \,\left\langle E_{1}\right\rangle \right)  }{\left\langle
A_{1}\right\rangle }\sigma_{1}^{2}\left(  \left\langle E_{1}\right\rangle
\right)  \ ,\label{eq:sig_E_ODE}%
\end{equation}
where $Z\left(  \left\langle E_{1}\right\rangle \right)  \equiv B_{11}/\left(
2A_{1}\right)  $. Solving the ordinary differential equation Eq.
(\ref{eq:sig_E_ODE}) and using Eq. (\ref{eq:full_relation}) we find for the
equilibrating systems that the variance is given by%
\begin{align}
\sigma_{1,eq}^{2}\left(  \left\langle E_{1}\right\rangle \right)  = &
\sigma_{1_{0}}^{2}\frac{A_{1}^{2}\left(  \left\langle E_{1}\right\rangle
\right)  }{A_{1}^{2}\left(  \left\langle E_{1}\right\rangle _{0}\right)
}+\nonumber\\
&  2A_{1}^{2}\left(  \left\langle E_{1}\right\rangle \right)  \int
_{\left\langle E_{1}\right\rangle _{0}}^{\left\langle E_{1}\right\rangle
}\frac{1}{A_{1}^{2}\left(  E^{\prime}\right)  \left(  \beta_{1}-\beta
_{2}\right)  }dE^{\prime}\ .\label{eq:equilibrium_approach}%
\end{align}
Here $\left\langle E_{1}\right\rangle _{0}$ and $\sigma_{1_{0}}^{2}$ are
$\left\langle E_{1}\right\rangle $ and $\sigma_{1}^{2}$ respectively at the
initial time. Recall that $E_{total}$ is held constant in this expression. It
is easy to extend these results when $E_{total}$ varies between experiments.
It is interesting to note that this expression is identical to that obtained
for a single driven isolated system \cite{NatPhys} when $\beta_{2}$ is set to
zero. This means that within this theory, driving a system is formally
equivalent to attaching it to a bath with infinite temperature. It is
straightforward to show, that when system 2 is a bath, so that $\beta_{2}$ can
be taken to be a constant, the width $\sigma_{1}^{2}$ approaches the
equilibrium value: $\left(  \partial_{E_{1}}\beta_{1}\right)  _{E_{eq}}%
^{-1}=k_{B}T^{2}C$, were $E_{eq}$ is the equilibrium value of $\left\langle
E_{1}\right\rangle $, and $C$ is the heat-capacity (see e.g., \cite{kardar}).
To see this, note that at equilibrium $A_{1}$ must vanish, and $\beta
_{1}=\beta_{2}$. Therefore the entire expression for $\sigma_{1}^{2}$ is
controlled by the final approach of $E$ to $E_{eq}$ where%
\begin{align*}
\beta_{1} &  \simeq\beta_{2}+\left(  \partial_{E_{1}}\beta_{2}\right)
_{E_{eq}}\left(  E-E_{eq}\right)  \ ,\\
A_{1}\left(  E^{\prime}\right)   &  =\left.  \frac{dA_{1}}{dE_{1}}\right\vert
_{E_{eq}}\left(  E-E_{eq}\right)  \ ,
\end{align*}
and the equilibrium expression follows. Note that away from the final
equilibration regime $A_{1}\left(  E\right)  $ need not be linear.

In the case of driven-dissipative systems (when system 2 is large), we obtain
for the variance%
\begin{equation}
\sigma_{1,dd}^{2}\left(  \left\langle E_{1}\right\rangle \right)
=\sigma_{1_{0}}^{2}\frac{A_{1}^{2}\left(  \left\langle E_{1}\right\rangle
\right)  }{A_{1}^{2}\left(  E_{1_{0}}\right)  }+2A_{1}^{2}\left(  \left\langle
E_{1}\right\rangle \right)  \int_{E_{1_{0}}}^{\left\langle E_{1}\right\rangle
}\frac{Z\left(  E^{\prime}\right)  }{A_{1}^{2}\left(  E^{\prime}\right)
}dE^{\prime}\ .\label{eq:steady_state_approach}%
\end{equation}
where $Z=\left[  1-\beta_{2}A_{F}/\left(  \beta_{1}A_{1}\right)  \right]
/\left(  \beta_{1}-\beta_{2}\right)  $. Eqs. (\ref{eq:equilibrium_approach})
and (\ref{eq:steady_state_approach}) are our main results for the approach to
steady-state. They predict the size of fluctuations in $E_{1}$ around its
average value. They depend only on the rates of energy injection into the
system $A_{F}$\ (which is zero for equilibrating systems) and the rate of
energy transfer to the bath $A_{2}$. In principle both these quantities can be
measured separately. $A_{F}$ can be measured by the rate of energy absorption
when system 1 is isolated, and $A_{2}$ in an equilibration experiment without
the drive. This is a consequence of our assumption of statistical independence
of the driving and the mechanism of interaction between the systems. In
addition we comment that Eqs. (\ref{eq:equilibrium_approach}) and
(\ref{eq:steady_state_approach}) imply that $\sigma_{1,eq}^{2}$ and
$\sigma_{1,dd}^{2}$ scale as $E\propto N$ when $A$ is a homogeneous function
of $N$ (e.g., extensive in $N$). This justifies self-consistently our
assumption on the narrowness of the distribution.

\emph{Steady-state fluctuations -} The framework described above can also be
used to study fluctuations in the steady-state of driven-dissipative systems,
specifically fluctuations of $E_{1}$ around $\left\langle E_{1}\right\rangle
$. At the steady-state the probability distribution $P_{s}\left(
E_{1}\right)  $ is independent of time. Using $\partial_{t}\left\langle
E_{1}\right\rangle =\left\langle A\right\rangle =A\left(  \left\langle
E_{1}\right\rangle \right)  $, and noting that at the steady-state $A\left(
\left\langle E_{1}\right\rangle \right)  $ must vanish, we expand $A_{1}$ and
$B_{11}$ to lowest order in $e_{1}\equiv E_{1}-E_{1}^{0}$%
\begin{equation}
A_{1}=-\frac{1}{\tau}e_{1}\ ,\ \ \ B_{11}=B_{s}\label{eq:As_Bs}%
\end{equation}
where $B_{s}$ and $\tau$ are constants. Equivalently, in this regime the
Fokker-Planck equation describes the Brownian motion of the energy in a
harmonic potential $\dot{e}_{1}=-e_{1}/\tau+\sqrt{B_{s}}\eta$, where the white
noise $\eta\left(  t\right)  $ satisfies $\left\langle \eta\left(  t\right)
\eta\left(  t^{\prime}\right)  \right\rangle =\delta\left(  t-t^{\prime
}\right)  $. $\tau$ is then interpreted as the relaxation time, as can be seen
from the two time correlation function%
\[
\left\langle e_{1}\left(  t_{1}\right)  e_{1}\left(  t_{2}\right)
\right\rangle =\frac{B_{s}\tau}{2}e^{-\left\vert t_{2}-t_{1}\right\vert /\tau
}\ .
\]
The variance of the energy fluctuations is given by $\sigma_{1}^{2}%
=\left\langle e_{1}\left(  t_{1}\right)  ^{2}\right\rangle =B_{s}\tau/2$.

When $A_{F}=0$, Eqs. (\ref{eq:full_relation}) and (\ref{eq:As_Bs}) imply
that$\ e_{1}=-\frac{\tau B_{s}}{2}\left(  \beta_{1}-\beta_{2}\right)  $. Then
expanding $\beta_{1}$ around $\beta_{2}$ as done above we find that
$\sigma_{1}^{2}=B_{s}\tau/2=-\left(  \partial_{E_{1}^{0}}\beta_{1}\right)
^{-1}$ which again reproduces the canonical distribution width. The present
derivation gives a dynamic interpretation to this formula.

When $A_{F}\neq0$ namely for a driven-dissipative system we find, using
$A_{1}\left(  E_{1}^{0}\right)  =0$ in Eq. (\ref{eq:full_relation}), and
$\sigma_{1}^{2}=B_{s}\tau/2$, that%
\begin{equation}
\tau A_{F}=\frac{\beta_{1}}{\beta_{2}}\left(  \beta_{2}-\beta_{1}\right)
\sigma_{1}^{2}\ .\label{eq:steady_state_tau}%
\end{equation}
This is our main result for the steady-state of driven-dissipative systems.
$A_{F}$ is the rate of energy injected to the system from the drive. In the
steady-state, this energy is then dissipated into the bath.\ This expression
therefore relates three central quantities characterizing the steady-state:
the size of the energy fluctuations $\sigma_{1}^{2}$, the rate of energy
dissipation $A_{F}$, and $\tau$ which is the relaxation time in the
steady-state \cite{2_baths_comment}.

\emph{MD\ Simulations - }Before proving the key relation Eq.
(\ref{eq:full_relation}), we illustrate our main results on a gas of
hard-sphere particles in a box, simulated by an event-driven molecular
dynamics simulation \cite{MDbook}. The gas is composed of $N_{1}$ particles of
mass $m_{1}$ and $N_{2}$ particles of mass $m_{2}$, all of equal size,
corresponding to systems 1 and 2 respectively. Although the entropy of the two
systems between collisions indeed factorizes, the collision process involves a
strong interaction, which changes the velocities of the particles by a
significant amount. A collision calculation shows that if the two masses are
very different, the energy transfer in each collision is small. In this case
energy transfer occurs over many collisions, fulfilling the assumption of
time-scale separation (see above). In what follows we take $m_{1}=10^{-4}$,
$m_{2}=1$. (Throughout we use arbitrary units). The box is a unit cube with
reflecting boundary-conditions, and the particles are taken to occupy a volume
fraction of $0.05$.

\begin{figure}
[ptb]
\begin{center}
\includegraphics[
height=1.9709in,
width=2.8781in
]%
{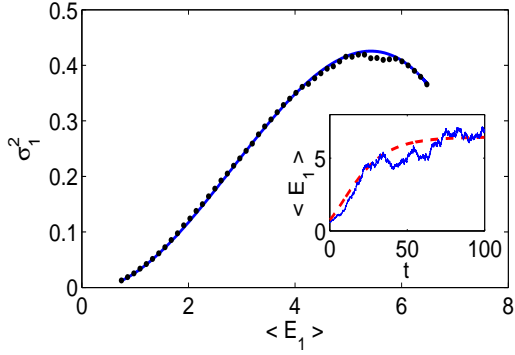}%
\caption{Equilibration of a system of 50 particles with two different masses.
Plotted are simulation results for $\left\langle E_{1}\right\rangle $ vs.
$\sigma_{1}^{2}$ (dots) compared to the theoretical prediction (solid line),
Eq. (\ref{eq:equilibrium_approach}). Inset: Example of $E_{1}\left(  t\right)
$ in a single run (solid line), and the average energy $\left\langle
E_{1}\right\rangle \left(  t\right)  $ (dashed line), used to calculate
$A_{1}$. }%
\label{fig:mdequil}%
\end{center}
\end{figure}

We first consider the approach to equilibrium of two systems in contact, Fig.
\ref{fig:settings}(a), to be compared with the predictions of Eq.
(\ref{eq:equilibrium_approach}). We take $N_{1}=30$ for the first systems and
$N_{2}=20$ for the second system. $N_{1},N_{2}$ are chosen to be relatively
small in order to test the theory on a mesoscopic system. The initial
velocities are sampled from a Maxwell-Boltzmann distribution with $\beta
_{1}=60$ and $\beta_{2}=3$, corresponding to average energies per particle of
$\left\langle E_{1}\right\rangle /N_{1}=0.025$ and $\left\langle
E_{2}\right\rangle /N_{2}=0.5$. We start all runs from a fixed total energy
$E_{total}=\left\langle E_{1}\right\rangle +\left\langle E_{2}\right\rangle $,
by preforming a (small) rescaling of the $m_{2}$-particles' velocities.
Gathering statistics over many runs, we calculate at each time the average
energy $\left\langle E_{1}\right\rangle \left(  t\right)  $ and the variance
$\sigma_{1}^{2}\left(  t\right)  $. The function $A_{1}\left(  \left\langle
E_{1}\right\rangle \right)  $ is obtained by plotting $A_{1}\left(  t\right)
=d\left\langle E_{1}\right\rangle /dt$ as a function of $\left\langle
E_{1}\right\rangle \left(  t\right)  $. Given $A_{1}\left(  \left\langle
E_{1}\right\rangle \right)  $ \cite{Max_model_comment}\ we use Eq.
(\ref{eq:equilibrium_approach}) to predict $\sigma_{1}^{2}\left(  \left\langle
E_{1}\right\rangle \right)  $, and find a good fit with the simulation
results, see Fig. \ref{fig:mdequil}.

\begin{figure}
[ptb]
\begin{center}
\includegraphics[
height=1.9709in,
width=2.8783in
]%
{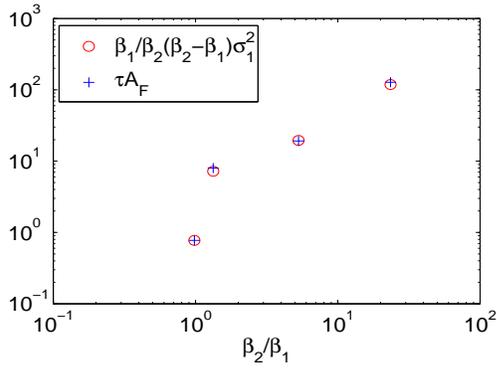}%
\caption{Test of Eq.\ (\ref{eq:steady_state_tau}) for a driven-dissipative
system. . }%
\label{fig:mdsteady_state}%
\end{center}
\end{figure}

We now turn to the driven-dissipative scenario, and test Eq.
(\ref{eq:steady_state_tau}) for the steady-state. We run simulations on a
system with $N_{1}=10$ and $N_{2}=50$ particles. System 1 is driven by
applying short impulses to the $m_{1}$-particles, changing their velocity by
$\Delta\mathbf{v}=\mathbf{F}\delta t/m_{1}$, where $\mathbf{F}$ is a constant
force and $\delta t$ the impulse duration. This impulse is applied at a
constant rate. In order to mimic the behavior of a very large system 2, the
velocity of the $m_{2}$ particles is changed upon reflection from the wall
\cite{MD_BCs}, so as to maintain a constant $\left\langle E_{2}\right\rangle
$.\ The quantities $\tau,A_{F},\beta_{1}$ and $\sigma_{1}^{2}$ are computed
from the numerics. The energy of system 2 is maintained at $\left\langle
E_{2}\right\rangle /N_{2}=1/2$, or $\beta_{M}=3$. Fig.
\ref{fig:mdsteady_state} shows the results obtained for the two sides of Eq.
(\ref{eq:steady_state_tau}) as a function of $\beta_{1}/\beta_{2}$ for
different strengths of the drive. Good agreement is found over a wide range of
drive strengths, and temperature differences. In this range, $A_{F}$ increases
by a factor of 1000, $\sigma_{1}^{2}$ by a factor of 17, and the relaxation
time $\tau$ decreases by a factor of 3.%

\emph{Derivation of the eq. (\ref{eq:full_relation}) -} We end by deriving Eq.
(\ref{eq:full_relation}).\ To do so we look at two times $t,t+\Delta t$\ in
which the driving protocol has returned to its original state, i.e. where
$H\left(  t\right)  =H\left(  t+\Delta t\right)  $, where $H$ is the
Hamiltonian of the combined system. As stated before, we assume that both
subsystems are relaxed in their respective energy shells, with energies
$E_{1}$ and $E_{2}$. In particular, this requires that $\Delta t\gg\tau_{R}$.
We denote the changes in $E_{1},E_{2}$ during the time interval $\Delta t$ by
$\Delta E_{1},\Delta E_{2}$ respectively, and define $\Delta E_{B}$ and
$\Delta E_{F}$ via
\[
\Delta E_{1}=\Delta E_{F}-\Delta E_{B}\ ,\ \Delta E_{2}=\Delta E_{B}\ .
\]
$\Delta E_{B}$ is the energy transferred from system 1 to system 2, and
$\Delta E_{F}$ is the work done on system 1 by the external drive.

Under these assumptions, Liouville's theorem or the unitarity of the dynamics,
together with micro-reversibility of the dynamics, imply a Crooks relation for
the combined isolated system \cite{crooks,isolated_crooks,QM_crooks}%
\begin{align*}
&  P_{E_{1},E_{2}}\left(  \Delta E_{1},\Delta E_{2}\right)  e^{S_{1}\left(
E_{1}\right)  +S_{2}\left(  E_{2}\right)  }\\
&  =\tilde{P}_{E_{1}+\Delta E_{1},E_{2}+\Delta E_{2}}\left(  -\Delta
E_{1},-\Delta E_{2}\right)  e^{S_{1}\left(  E_{1}+\Delta E_{1}\right)
+S_{2}\left(  E_{2}+\Delta E_{2}\right)  }\ ,
\end{align*}
where $P_{E_{1},E_{2}}\left(  \Delta E_{1},\Delta E_{2}\right)  $ is the
probability of a transition $\left(  E_{1},E_{2}\right)  \rightarrow\left(
E_{1}+\Delta E_{1},E_{2}+\Delta E_{2}\right)  $, and $\tilde{P}$ is defined
similarly, only with respect to the the reversed protocol, defined by the
dynamics generated by the time-reversed Hamiltonian, $\tilde{H}\left(
t^{\prime}\right)  =H\left(  t+\Delta t-t^{\prime}\right)  $.

We assume that: (1) $\Delta E_{1},\Delta E_{2}$ are small so that the
transition probabilities depend weakly on the initial energies $\tilde
{P}_{E_{1}+\Delta E_{1},E_{2}+\Delta E_{2}}=\tilde{P}_{E_{1},E_{2}}$, with
corrections which are of order $N^{-1}$ \cite{NatPhys}. (2) $\Delta E_{B}$ and
$\Delta E_{F}$ are statistically independent quantities. This happens when the
interaction with the bath is independent from the driving process, e.g., when
the drive and interaction processes act on different modes, on different parts
of the system, at different times, etc.. These imply that%
\begin{align*}
&  P^{F}\left(  \Delta E_{F},0\right)  P^{B}\left(  -\Delta E_{B},\Delta
E_{B}\right)  e^{-\beta_{1}\delta E_{F}-\left(  \beta_{2}-\beta_{1}\right)
\delta E_{B}}\\
&  =\tilde{P}^{F}\left(  \Delta E_{F},0\right)  \tilde{P}^{B}\left(  -\Delta
E_{B},\Delta E_{B}\right)  \ .
\end{align*}
Integrating over $\Delta E_{F},\Delta E_{B}$ gives a Jarzynski relation
$\left\langle e^{-\beta_{1}\Delta E_{F}}\right\rangle \left\langle e^{-\left(
\beta_{2}-\beta_{1}\right)  \Delta E_{B}}\right\rangle =1$. Rearranging the
equation and again integrating gives $\left\langle e^{-\beta_{1}\Delta E_{F}%
}\right\rangle =\left\langle e^{-\left(  \beta_{2}-\beta_{1}\right)  \Delta
E_{B}}\right\rangle $ yielding $\left\langle e^{-\beta_{1}\Delta E_{F}%
}\right\rangle =\left\langle e^{-\left(  \beta_{2}-\beta_{1}\right)  \Delta
E_{B}}\right\rangle =1$. The second relation is the exchange fluctuation
relation \cite{fluct_heat_exchange,QM_fluct_relation_review}; The first is a
variant of the Jarzynski relation for isolated systems \cite{NatPhys}.
Expanding both to second order we find%
\[
2\left\langle \Delta E_{F}\right\rangle =\beta_{1}\left\langle \Delta
E_{F}^{2}\right\rangle \ ,\ 2\left\langle \Delta E_{B}\right\rangle =\left(
\beta_{2}-\beta_{1}\right)  \left\langle \Delta E_{B}^{2}\right\rangle \ .
\]
From independence it follows that $\left\langle \Delta E_{F}\Delta
E_{1}\right\rangle =\left\langle \Delta E_{F}^{2}\right\rangle $, or
$\left\langle \Delta E_{B}^{2}\right\rangle =\left\langle \Delta E_{1}%
^{2}\right\rangle -\left\langle \Delta E_{F}^{2}\right\rangle $. The second
equation becomes $2\left\langle \Delta E_{B}\right\rangle =\left(  \beta
_{2}-\beta_{1}\right)  \left(  \left\langle \Delta E_{1}^{2}\right\rangle
-\left\langle \Delta E_{F}^{2}\right\rangle \right)  $, so that
\[
2\left\langle \Delta E_{1}\right\rangle -2\beta_{2}/\beta_{1}\left\langle
\Delta E_{F}\right\rangle =\left(  \beta_{1}-\beta_{2}\right)  \left\langle
\Delta E_{1}^{2}\right\rangle \ .
\]
After dividing by $\Delta t$ and using the definitions of the $A_{i},B_{i}$
quantities, yields Eq. (\ref{eq:full_relation}).

\emph{Acknowledgments -} We are grateful to Luca D'Alessio, Dov Levine, Daniel
Podolsky and Anatoli Polkovnikov for many useful discussions and comments. The
work was supported by a BSF grant. YK thanks G. M. Schutz.


\begin{thebibliography}{99}
\bibitem {cumulant_comment}For finite $\Delta t$ one should in principle
consider cumulents. Here, as $\Delta t$ is short, the distinction disappears.
\bibitem {2_baths_comment}A similar derivation can be carried out for a system
coupled to two baths at temperatures $\beta_{2}$ and $\beta_{3}$, instead of
one bath and a drive. One then obtains the for the steady-state $A_{F}%
\tau=\left(  \beta_{1}-\beta_{2}\right)  \left(  \beta_{3}-\beta_{1}\right)
\left(  \beta_{3}-\beta_{2}\right)  ^{-1}\sigma_{1}^{2}$.
\bibitem {Max_model_comment}Incidentally, $\left\langle E_{1}\right\rangle
\left(  t\right)  $ is found to fit very well to a function of the form
$E=E_{fin}\tanh^{2}\left(  a+bt\right)  $, or $A_{1}\propto\left\langle
E\right\rangle ^{1/2}\left(  \left\langle E\right\rangle -E_{fin}\right)  $,
allowing us to write a closed analytical expression for $\sigma_{1}^{2}\left(
\left\langle E\right\rangle \right)  $. This can be understood by calculating
the rate of energy transfer, taking the collision rate to depend only on the
velocity of the lighter, and hence faster, $m_{1}$ particles.
\bibitem {QM_crooks}In the quantum mechanical setting we assume that the
initial and final density matrices are diagonal, as in \cite{NatPhys}, and
that there is no significant entangelment between the two systems.
\end{thebibliography}
\end{document}